\documentclass[11pt]{article}

\usepackage[dvips]{color}
\usepackage{epsfig}
\usepackage{amsmath}
\usepackage{graphicx}
\usepackage{amsfonts}
\usepackage{graphicx}% Include figure files
\usepackage{dcolumn}% Align table columns on decimal point
\usepackage{bm}% bold math
\usepackage{float}

\date{}
\begin{document}

\title{\bf High temperature dimensional reduction in Snyder space}

\author{K. Nozari\thanks{e-mail: knozari@umz.ac.ir},\hspace{.3cm}
V. Hosseinzadeh\thanks{e-mail: v.hosseinzadeh@stu.umz.ac.ir}\hspace{.3cm}
and\hspace{.3cm}M. A. Gorji\thanks{e-mail: m.gorji@stu.umz.ac.ir}
\\\\{\small {\it Department of Physics, Faculty of Basic Sciences,
University of Mazandaran,}}\\{\small {\it P.O. Box 47416-95447,
Babolsar, Iran}}}\maketitle

\begin{abstract}
In this paper, we formulate the statistical mechanics in Snyder space that
supports the existence of a minimal length scale. We obtain the corresponding
invariant Liouville volume which properly determines the number of microstates
in the semiclassical regime. The results show that the number of accessible
microstates drastically reduces at the high energy regime such that there is
only one degree of freedom for a particle. Using the Liouville volume, we
obtain the deformed partition function and we then study the thermodynamical
properties of the ideal gas in this setup. Invoking the equipartition theorem,
we show that $2/3$ of the degrees of freedom freeze at the high temperature
regime when the thermal de Broglie wavelength becomes of the order of the
Planck length. This reduction of the number of degrees of freedom suggests an
effective dimensional reduction of the space from $3$ to $1$ at the Planck
scale.
\begin{description}
\item[PACS numbers]
04.60.Bc
\item[Key Words]
Quantum Gravity Phenomenology, Thermodynamics
\end{description}
\end{abstract}
\section{Introduction}
While general relativity and quantum mechanics are successful in
their applicability domains, it seems that there is a fundamental
incompatibility between them in order to find the so-called quantum
theory of gravity. Such a theory, not completely formulated so far,
would reasonably describe the structure of spacetime at the Planck
scale where both of the gravitational and quantum mechanical effects
become important. Despite the fact that there is no unique approach
to quantum gravity, existence of a universal minimum measurable
length, preferably of the order of the Planck length
$l_{_{Pl}}\sim\,10^{-33}\,m$, is a common feature of quantum gravity
candidates such as string theory and loop quantum gravity
\cite{String,LQG}. It is then widely believed that a
non-gravitational theory which includes a universal minimal length
scale would appear at the flat limit of quantum gravity. Therefore,
many attempts have been done in order to take into account a minimal
length scale in the well-known non-gravitational theories such as
quantum mechanics and special relativity. The generalized
uncertainty principle is investigated in the context of the string
theory that supports the existence of a minimal length as a nonzero
uncertainty in position measurement \cite{GUP}. Quantum field
theories turn out to be naturally ultraviolet-regularized in this
setup \cite{GUP-QFT}. Inspired by the seminal work of Snyder in 1947
who was formulated a Lorentz-invariant noncommutative spacetime
\cite{Snyder}, a phase space with noncanonical symplectic structure
is formulated in the non-relativistic limit \cite{Mignemi}. At the
quantum level, this deformed phase space leads to the modified
uncertainty relation which is very similar to one arises from the
string theory motivations \cite{Mignemi}. Furthermore, recently,
polymer quantum mechanics is suggested in the symmetric sector of
loop quantum gravity which supports the existence of a minimal
length scale known as the polymer length scale \cite{QPR}. Also, the
doubly special relativity theories are investigated in order to take
into account a minimal observer-independent length scale in special
relativity \cite{DSR}. Appearance of curved energy-momentum space is
the direct consequence of the doubly special relativity theories
\cite{Curve-M} and, interestingly, the Snyder noncommutative
spacetime could be also realized in this setup by a relevant gauge
fixing process \cite{DSR-Snyder}.

Apart from the details of the above mentioned phenomenological
models as a flat limit for quantum gravity, all of them suggest the
deformation to the density of states at the high energy regime which
in turn leads to the nonuniform measure over the set of microstates.
Indeed, in these setups, the number of accessible microstates will
be reduced at the high energy regime due to the existence of a
minimal length as an ultraviolet cutoff for the system under
consideration. Reduction of the number of degrees of freedom,
however, immediately suggests an effective dimensional reduction of
the space. This consequence seems to be a general feature of quantum
gravity which may also open new window for the statistical origin of
black holes thermodynamics \cite{QG-DR}. Thermodynamics of black
holes are widely studied in the frameworks of phenomenological
quantum gravity models such as noncommutative space \cite{BH-NC},
generalized uncertainty principle \cite{BH-GUP} and polymer
quantization scenario \cite{PL-BH}. The reduction of the number of
accessible microstates due to the universal quantum gravitational
effects would also significantly change the thermodynamical
properties of any physical system at the high temperature regime.
Therefore, quantum gravity effects on the thermodynamics of various
statistical systems are widely studied in different contexts
\cite{QGP-THR}. For the special case of the ideal gas, it is natural
to expect that the quantum gravity effects would become important at
the high temperature regime, when the corresponding thermal de
Broglie wavelength $\lambda=\sqrt{\frac{2\pi}{mT}} \,\hbar$ becomes
of the order of the Planck length $l_{_{\rm Pl}}=\sqrt{\hbar{G} }$,
where $m$ is the particles' mass and $T$ denotes the
temperature\footnote{We work in units $k_{_B}=1=c$, where $k_{_B}$
and $c$ are the Boltzmann constant and speed of light in vacuum
respectively.}. The associated thermodynamical properties then will
be significantly modified in this regime. Thermodynamics of the
ideal quantum gases in noncommutative space are studied in Refs.
\cite{NC-THR-IG,Snyder-Ther} and for the case of the effects that
arise from the generalized uncertainty principle see Ref.
\cite{GUP-THR-IG}. Thermodynamical properties of the ideal gas in
polymerized phase space, as a classical limit of a polymer quantum
mechanics, is also studied in Refs. \cite{PL-THR-IG,Husain,PL-DOS}.
For the case of the relativistic ideal gases in doubly special
relativity framework see Refs. \cite{Glikman,DSR-THR-IG}. Motivated
by the above stated issues, in this paper we study the
thermodynamical properties of the ideal gas in Snyder space.

The structure of the paper is as follows: In section 2, the
statistical mechanics in the Snyder space is formulated and the
corresponding partition function is found. Using the partition
function, thermodynamics of the ideal gas is studied in section 3.
Section 4 is devoted to the summary and conclusions.

\section{Statistical Mechanics in Snyder Space}
The kinematics and dynamics of a classical system on the
phase space provide a suitable framework for formulating the
statistical mechanics in the semiclassical regime. The key
quantity is the Liouville volume that determines the density
of states from which all the thermodynamical properties of
a system could be achieved. In this section, using the
symplectic geometry, we formulate the statistical mechanics
in Snyder space.

\subsection{Kinematics and Dynamics}
Inspired by the seminal work of Snyder on noncommutative spacetime
\cite{Snyder}, the associated deformed phase space is formulated
which also supports the existence of a minimal length
\cite{Mignemi}. A phase space naturally admits symplectic structure
and therefore is a symplectic manifold. Suppose that
$(\Gamma,\omega)$ to be a Snyder-deformed phase space with $\omega$
as the associated symplectic structure which is a closed
nondegenerate $2$-form on $\Gamma$. The local form of the symplectic
structure in Snyder model is given by \cite{Stern}
\begin{eqnarray}\label{Snyder-symplectic}
\omega=dq^i\wedge dp_i-\frac{1}{2}\,d(q^ip_i)\,{\wedge}\,
d\ln\left[1+\beta^2p^2\right]\,,
\end{eqnarray}
where $q^i$ and $p_i$ are the position and momentum coordinates of a
particle with $i,j=1,...,3$ and $p^2= \delta^{ij}p_i p_j$. $\beta$
is the deformation parameter with dimension of length which is
usually taken to be of the order of the Planck length as
$\beta=\beta_0\, l_{_{\rm Pl}}$, where $\beta_0={\mathcal O}(1)$ is
the dimensionless numerical constant that should be fixed only with
experiment \cite{QGExperiment}. Taking the low energy limit
$\beta\rightarrow\,0$ in the relation (\ref{Snyder-symplectic}), the
standard well-known canonical form of the symplectic structure could
be recovered.

Since the symplectic structure is nondegenerate by
definition, one can assign a unique vector field
${\mathbf x}_{_f}$ to any function $f$ on $\Gamma$ as
$\omega\,({\mathbf x}_{_f})=df$. The Poisson bracket for
two real-valued functions is defined as
\begin{equation}\label{PB1}
\{f,\,g\}=\omega({\mathbf x}_{_f},{\mathbf x}_{_g})\,.
\end{equation}
From the above definition, it is straightforward to show
that the symplectic structure (\ref{Snyder-symplectic})
generates the following noncanonical Poisson algebra
\begin{eqnarray}\label{Snyder A}
\{q^i,q^j\}=\beta^2J^{ij},\hspace{.7cm}\{q^i,p_j\}=
\delta^i_j+\beta^2\,p^ip_j,\hspace{.7cm}\{p_i,p_j\}=0,
\end{eqnarray}
where $J_{ij}=q_ip_j-q_jp_i$ is the generator of the
rotation group in three dimensions with $q_i=\delta_{ij}
q^j$. The symplectic structure (\ref{Snyder-symplectic})
or equivalently the Poisson algebra (\ref{Snyder A})
properly defines the kinematics of the phase space
$\Gamma$ in Snyder model.

The dynamics of the system will be determined by specifying a
Hamiltonian function $H$ as the generator of time evolution of the
system. The Hamiltonian system on the phase space is then defined by
the triplet $(\Gamma, \omega,H)$ and the dynamical evolution of the
system is governed by the equation
\begin{equation}\label{dynamic}
\omega\,({\mathbf x}_{_H})=dH\,,
\end{equation}
where ${\mathbf x}_{_H}$ is the Hamiltonian vector field
and it's integral curves are nothing but the Hamilton's
equations in this setup (see Ref. \cite{Stern} for more
details).

Furthermore, the natural volume on the phase space is
the Liouville volume which for a $2n$-dimensional phase
space is defined as
\begin{eqnarray}\label{Vol}
\omega^{n}=\frac{1}{n!}\,\omega\,\wedge...\wedge\,\omega
\hspace{1cm}(n\,\,\mbox{times})\,.
\end{eqnarray}
The Liouville volume for a particle in Snyder-deformed
phase space then could be easily obtained by
substituting the symplectic structure
(\ref{Snyder-symplectic}) into the definition
(\ref{Vol}) which gives
\begin{eqnarray}\label{Snyder-measure}
\omega^3=dq^1\wedge dq^2\wedge dq^3\,\wedge\,\frac{dp_1
\,\wedge{dp}_2\,\wedge{dp}_3}{\big(1+\beta^2p^2\big)}\,.
\end{eqnarray}
The phase space (Liouville) volume determines the density
of states and then the number of accessible microstates
for a statistical system. It is important to check
the verification of the Liouville theorem for the Snyder
measure (\ref{Snyder-measure}) in order to formulate the
statistical mechanics in Snyder-deformed phase space. The
Liouville theorem states that the Liouville volume is
invariant under the time evolution of the system
\begin{equation}\label{Liouvillee}
\frac{d\omega^n}{dt}=\frac{\partial\omega^n}{\partial t}
+{\mathcal L}_{{\mathbf x}_{_H}}\omega^n=0\,,
\end{equation}
where ${\mathcal L}_{{\bf x}_{_H}}$ denotes the Lie
derivative with respect to ${\bf x}_{_H}$. The relation
(\ref{Liouvillee}) can be traced back to the facts that
$\omega^{n}$ is not explicitly time-dependent,
${\mathcal{L}}_{{\mathbf x}_{_H}}\omega^{n}=n(
{\mathcal{L}}_{{\mathbf x}_{_H}}\omega)\wedge\omega^{
n-1}$ and ${\mathcal{L}}_{{\mathbf x}_{_H}}\omega=d(
\omega\,({{\mathbf x}_{_H}}))+(d\omega)\,({
{\mathbf x}_{_H}})=0$, where we have used the equation
(\ref{dynamic}) as $d(\omega\,({{\mathbf x}_{_H}}))=
d^2H=0$ and the closure of the symplectic structure
$d\omega=0$. The result (\ref{Liouvillee}) for the
Snyder measure (\ref{Snyder-measure}) is essential
for us to formulate the statistical mechanics in
Snyder space.

\subsection{Number of Microstates}
Before obtaining the partition function, from which
all the thermodynamical properties of a system can
be obtained, we first make a qualitative discussion
about the number of microstates in Snyder-deformed
phase space regardless of the ensemble density and
the Hamiltonian function.

The number of microstates for a single-particle
phase space is given by
\begin{equation}\label{Nm}
N_m=\frac{1}{h^3}\int_{p\leq{p_{\ast}}}\,\omega^3\,,
\end{equation}
where we have considered a region of phase space in which the
condition $p\leq{p_{\ast}}$, with $p=\sqrt{\delta^{ ij}p_ip_j}$ is
satisfied in order to have a finite number of microstates.

\begin{figure}
\flushleft\leftskip+15em{\includegraphics[width=3in]{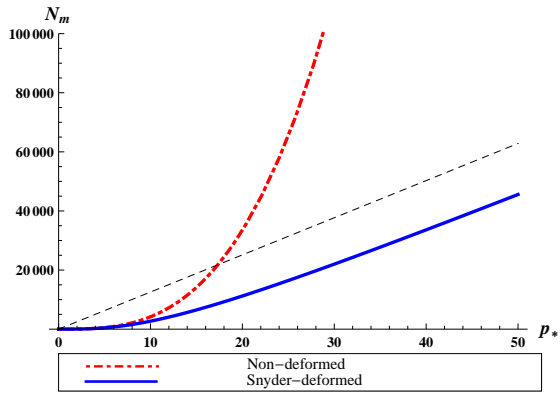}}
\hspace{3cm}\caption{\label{fig:0} The number of microstates versus
the momentum $p_\ast$ for a single-particle phase space. The blue
solid and red dot-dashed lines represent the number of microstates
in the Snyder-deformed and non-deformed phase spaces respectively.
Clearly, these two curves coincide at the low energy regime
$\beta{p_\ast}\ll{1}$ ($p_\ast\ll{ p_{_{\rm Pl}}}$) and the
deviation arises at the high energy regime $\beta{p_\ast}\sim{1}$
($p_\ast\sim{ p_{_{\rm Pl}}}$). The number of microstates in the
Snyder-deformed case effectively behaves like a one-dimensional
single-particle phase space (the black dashed line) which shows that
two degrees of freedom freeze at the high energy regime due to the
quantum gravity (minimal length) effects. This result suggests an
effective dimensional reduction of the space from $3$ to $1$
dimension at the high energy regime. The figure is plotted for
$\beta=\frac{\beta_0}{p_{_{\rm Pl}} }=0.1$ with $\beta_0=1$ and
$p_{_{\rm Pl}}=10$.}
\end{figure}

The quantity (\ref{Nm}) for a system with non-deformed
phase space is $N_m=({4\pi{V}}/{h^3})\int_0^{p_\ast}
p^2dp=({4\pi{V}}/{3h^3})\,p_\ast^3$, where $V$ is the
spatial volume of the system under consideration. For
the case of the Snyder-deformed phase space,
substituting the Liouville volume (\ref{Snyder-measure})
into the relation (\ref{Nm}), it works out as
\begin{equation}\label{Snyder-Nm}
N_m=\frac{4\pi{V}}{h^3}\int_0^{p_\ast}\frac{p^2dp}{
\big(1+\beta^2p^2\big)}=\frac{4\pi{V}}{h^3\beta^3}
\Big(\beta{p_\ast}-\arctan[\beta{p_\ast}]\Big)\,.
\end{equation}
For the low energy regime with $\beta{p_\ast}\ll{1}$ (
$p_\ast\ll{p_{_{\rm Pl}}}$), the relation (\ref{Snyder-Nm}) for the
number of microstates behaves as $N_m\sim\,p_\ast^3$ which shows
that it correctly coincides with the standard result of the
non-deformed case in this regime. For the high energy regime
$\beta{p_\ast}\sim{1}$ ($p_\ast\sim{p_{_{\rm Pl }}}$), however, the
number of microstates (\ref{Snyder-Nm}) behaves linearly with
$p_\ast$ as $N_m\sim{p_\ast}$ (see also figure \ref{fig:0}, where
the number of microstates (\ref{Snyder-Nm}) versus $p_\ast$ is
plotted and also it is compared with the standard non-deformed
case). This result shows that, at the high energy regime when the
quantum gravity (minimal length) effects dominate, the number of
microstates will be drastically decreased. Since the number of
microstates for a standard one-dimensional particle (with two
dimensional phase space) behaves linearly with the momentum, one
could conclude that two degrees of freedom will be frozen at the
high energy regime and therefore a particle has only one degree of
freedom in this regime. This result suggests an effective
dimensional reduction of the space from $3$ to $1$ dimension at the
high energy regime in Snyder model. Although this result is
qualitatively obtained here for a general system (without specifying
a Hamiltonian function and ensemble density), we will explicitly
justify this result in the next section for the particular case of
an ideal gas in canonical ensemble in the light of the well-known
equipartition theorem of energy.

\subsection{Partition Function}
In order to study statistical mechanics in this framework, we
generalize this setup to a many-particle system. Consider a
statistical system consisting of $N$ particles. The corresponding
kinematical phase space can be obtained by the direct coupling of
the single-particle phase spaces as
\begin{eqnarray}\label{Gamma-tot}
\Gamma_{\rm tot}=\Gamma_1\times...\times\Gamma_N\,,\hspace{1.5cm}
\omega_{\rm tot}=\sum_{\alpha=1}^N{\omega}_{\alpha}\,,
\end{eqnarray}
where $\omega_\alpha$ is the symplectic structure on the phase space
of the $\alpha$-th particle, $\Gamma_\alpha$. Substituting the
symplectic structure (\ref{Gamma-tot}) into the definition
(\ref{Vol}), the corresponding $6N$-dimensional Liouville volume
will be
\begin{align}\label{Vol-tot}
\omega^{3N}=\frac{1}{(3N)!}\,\bigg(\sum_{\alpha=1}^N{\omega}_{
\alpha}\bigg){\wedge\,.\,.\,.\,\wedge}\bigg(\sum_{\alpha=1}^N
{\omega}_{\alpha}\bigg)=\omega_1^3\,{\wedge\,...\,\wedge}\,
\omega_N^3\,,
\end{align}
where $\omega_\alpha^3$ is the six-dimensional Liouville volume
corresponding to the $\alpha$-th particle phase space and we have
also used the fact that $\omega^i_\alpha=0$ for $i>3$, with
$\omega^i_\alpha$ being the $i$-th component of the $\alpha$-th
particle's Liouville volume. The quantum gravity parameter $\beta$
is universal and it will be the same for all the particles.
Therefore, the symplectic structure for all of the particles in
Snyder-deformed phase space is given by (\ref{Snyder-symplectic})
and the Liouville volume is then given by (\ref{Snyder-measure}).
Substituting the Liouville volume (\ref{Snyder-measure}) for all of
the particles in the relation (\ref{Vol-tot}) gives
\begin{align}\label{Snyder-Vol}
\omega^{3N}=\frac{d^{3N}q\,d^{3N}p}{\big(1+\beta^2p^2\big)^N}\,,
\end{align}
where clearly the standard volume $d^{3N}q\,d^{3N}p$ for the
non-deformed $6N$-dimensional phase space could be recovered in the
low energy limit $\beta\rightarrow\,0$. While the non-deformed phase
space volume assigns a uniform probability distribution over the set
of microstates, the Snyder-deformed phase space volume
(\ref{Snyder-Vol}) assigns a nonuniform probability distribution at
the high energy regime such that the microstates with higher energy
are less probable. More precisely, in the absence of any extra
information for the system, the Laplace principle of indifference
states that all the microstates are equally likely \cite{Laplace}.
However, in the presence of a minimal length, as an extra
information for the system, the sufficient condition for the
Laplace's indifference principle is no longer satisfied in the
Snyder model.

The Liouville volume (\ref{Snyder-Vol}) is invariant
under the time evolution of the system on $\Gamma_{
\rm tot}$ which can be easily deduced from the relation
(\ref{Liouvillee}). Thus the Liouville theorem is
satisfied on the phase space of the $N$-particle system
$\Gamma_{\rm tot}$ which allows us to study the statistical
mechanics in this setup. The canonical partition function
for a system at the temperature $T$ in Snyder model then
will be
\begin{eqnarray}\label{pf-tot}
{\mathcal Z}_N=\frac{1}{h^{3N}N!}\int_{\Gamma_{\rm tot}}
\omega^{3N}\,\exp\big[-H/T\big]\,,
\end{eqnarray}
where the Gibbs factor $1/N!$ is also considered. For a
system in which the particles do not interact, the total
Hamiltonian function could be decomposed as $H_{\rm tot}
=\sum_{\alpha=1}^N H_\alpha$ and the partition function
(\ref{pf-tot}) simplifies to
\begin{eqnarray}\label{Snyder-pf-tot}
{\mathcal Z}_N=\frac{{\mathcal Z}_1^N}{N!}\,,
\end{eqnarray}
where we have defined the single-particle partition
function as
\begin{equation}\label{Snyder-pf}
{\mathcal Z}_1=\frac{1}{h^3}\int_{\Gamma}\omega^3\exp\big[
-H/T\big]\,=\frac{1}{h^3}\int{d^3q}\int\frac{d^3p}{\big(1+
\beta^2p^2\big)}\,\exp\big[-H(q,p)/T\big].
\end{equation}
Having the partition function (\ref{Snyder-pf-tot}) in
hand, one could easily study the thermodynamical
properties of the statistical systems in Snyder space.

\section{Thermodynamics of Ideal Gas}
In this section we study the thermodynamical properties of the ideal
gas by means of the results obtained in pervious section.

By considering an ideal gas consisting of $N$ noninteracting
particles confined in volume $V$ at the temperature $T$, the
corresponding single-particle partition function can be easily
obtained from the relation (\ref{Snyder-pf}) as
\begin{eqnarray}\label{pf0}
{\mathcal Z}_1[V,T]=\frac{4\pi{V}}{h^3}\int_{0}^{\infty}
\frac{\exp\big[-\frac{p^2}{2mT}\big]p^2dp}{\big(1+\beta^2
p^2\big)}=\frac{V\pi^{\frac{3}{2}}}{h^3\beta^3}
\Bigg(\sqrt{2mT}\beta-\sqrt{\pi}\,\mbox{erfc}\bigg[\frac{
1}{\sqrt{2mT}\beta}\bigg]\exp\bigg[\frac{1}{2mT\beta^2}
\bigg]\Bigg)\,,
\end{eqnarray}
where $\text{erfc}[x]=(2/\sqrt{\pi})\int_{x}^{\infty}e^{
-t^2}dt$ is the complementary error function and also we
have substituted $H=p^2/2m$ for the Hamiltonian function.
To understand the qualitative behavior of the partition
function (\ref{pf0}), it is useful to rewrite the
partition function in terms of the thermal wave length
$\lambda=\frac{h}{\sqrt{2\pi{m}T}}$ as
\begin{equation}\label{pf}
{\mathcal Z}_1[V,\lambda]=\frac{V}{\lambda_{_{\rm Pl}}^3}
\Bigg(\frac{\lambda_{_{Pl}}}{\lambda}-\sqrt{\pi}\,
\mbox{erfc}\bigg[\frac{\lambda}{\lambda_{_{\rm Pl}}}\bigg]
\exp\bigg[\frac{\lambda^2}{\lambda_{_{\rm Pl}}^2}\bigg]
\Bigg)\,,
\end{equation}
where we have substituted $\beta=\beta_0\,l_{_{\rm Pl
}}=\frac{\beta_0}{T_{_{\rm Pl}}}$ and also we have
defined {\it Planck scale thermal de Broglie
wavelength}
\begin{equation}\label{lambda-p}
\lambda_{_{\rm Pl}}=(\sqrt{2}\beta_0)\times\frac{h}{\sqrt{
2\pi{m_{_{\rm Pl}}}T_{_{\rm Pl}}}}=2\sqrt{\pi}\beta_0\,
l_{_{\rm Pl}}\,,
\end{equation}
where as we have mentioned before $\beta_0={\mathcal O }(1)$ should
be fixed by experiment \cite{QGExperiment}. Expanding partition
function (\ref{pf}) for both the low and high temperature regimes
gives
\begin{equation}\label{pf0-Exp}
{\mathcal Z}_1[V,\lambda]=
\left\{
  \begin{array}{ll}
    \frac{V}{\lambda^3} & \hspace{.5cm}\lambda
    \gg\lambda_{_{\rm Pl}},\\\\
    \frac{V}{\lambda\lambda_{_{\rm Pl}}^2} &
    \hspace{.5cm}\lambda\sim\lambda_{_{\rm Pl}}.
  \end{array}
\right.
\end{equation}
As it is clear from the above relation, quantum gravity (minimal
length) effects are negligible at the low temperature limit
$\lambda\gg\lambda_{_{ \rm Pl}}\,(T\ll{T}_{_{\rm Pl}})$ and the
standard result for the partition function of the ideal gas is
recovered. Interestingly, the high-temperature behavior of the
partition function (\ref{pf0-Exp}) shows that two degrees of freedom
will be frozen at the Planck scale $\lambda\sim \lambda_{_{\rm Pl}}$
and there is only one degree of freedom for a particle in this
regime. This feature, as we will show, leads to the effective
reduction of the dimension of space at the high temperature regime.

From the relation (\ref{Snyder-pf-tot}), the total
partition function for the ideal gas in Snyder
space will be
\begin{equation}\label{pf-total}
{\mathcal Z}_N[V,\lambda]=\frac{\big(V/\lambda_{_{
\rm Pl}}^3\big)^N}{N!}\Bigg(\frac{\lambda_{_{\rm Pl
}}}{\lambda}-\sqrt{\pi}\,\mbox{erfc}\bigg[\frac{
\lambda}{\lambda_{_{\rm Pl}}}\bigg]\exp\bigg[\frac{
\lambda^2}{\lambda_{_{\rm Pl}}^2}\bigg]\Bigg)^N\,,
\end{equation}
from which all of the thermodynamical quantities could be derived
through the standard definitions.
\subsection{Internal Energy and Specific Heat}
The Helmholtz free energy $F$ is defined as
\begin{eqnarray}\label{Helmholtz}
F=-T\ln\big[{\mathcal Z}_{N}[V,\lambda]\big]=-NT
\left\{1+\ln\Bigg[\frac{V}{N\lambda_{_{\rm Pl}}^3
}\Bigg(\frac{\lambda_{_{\rm Pl}}}{\lambda}-\sqrt{
\pi}\,\mbox{erfc}\bigg[\frac{\lambda}{\lambda_{_{
\rm Pl}}}\bigg]\exp\bigg[\frac{\lambda^2}{
\lambda_{_{\rm Pl}}^2}\bigg]\Bigg)\Bigg]\right\}\,,
\end{eqnarray}
where we have used the Stirling's formula $\ln[N!]
\approx N\ln[N]-N$ for large $N$.

\begin{figure}
\flushleft\leftskip+15em{\includegraphics[width=3in]{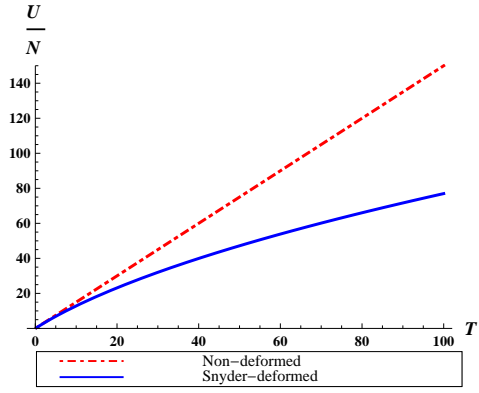}}
\hspace{3cm}\caption{\label{fig:1} Internal energy versus
temperature. The blue solid and red dot-dashed lines represent the
internal energy in Snyder-deformed and non-deformed phase spaces
respectively. As the temperature increases, the quantum gravity
(minimal length) effects become more and more appreciable.}
\end{figure}

\begin{figure}
\flushleft\leftskip+15em{\includegraphics[width=3in]{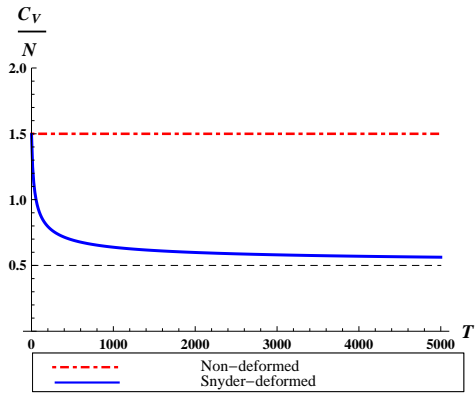}}
\hspace{3cm}\caption{\label{fig:2} Specific heat versus the
temperature. The blue solid and red dot-dashed lines represent the
specific heat in Snyder-deformed and non-deformed phase spaces
respectively. While the specific heat is independent of the
temperature as $\frac{3}{2}N$ for the case of standard ideal gas, it
becomes temperature-dependent at the high energy regime in Snyder
model. It asymptotically leads to the value $\frac{1 }{2}N$ at the
very high temperature regime which signals the effective dimensional
reduction of space from $3$ to $1$ dimension in this setup. It is
also clear from the figure that the specific heat is bounded as
$\frac{1}{2}\leq\frac{C_{_V}}{N}\leq\frac{ 3}{2}$ in Snyder model.}
\end{figure}

From the modified Helmholtz free energy
(\ref{Helmholtz}), the internal energy for the
ideal gas gets modified as
\begin{eqnarray}\label{energy}
U=-T^2\left(\frac{\partial}{\partial T}\bigg(
\frac{F}{T}\bigg)\right)_{N,V}=NT\left\{\Bigg(2-
\sqrt{\pi}\,\bigg(\frac{\lambda}{\lambda_{_{\rm
Pl}}}\bigg)\mbox{erfc}\bigg[\frac{\lambda}{
\lambda_{_{\rm Pl}}}\bigg]\exp\bigg[\frac{
\lambda^2}{\lambda_{_{\rm Pl}}^2}\bigg]\Bigg)^{
-1}-\frac{\lambda^2}{\lambda_{_{\rm Pl}}^2}
\right\}.
\end{eqnarray}
The internal energy versus the temperature is plotted in figure
\ref{fig:1}. As it is clear from the figure, while the internal
energy of the standard ideal gas increases linearly with
the temperature as $U_0=\frac{3}{2}NT$, which is shown by the
red dot-dashed line in the figure, the internal energy increases
with a decreasing rate at the high temperature regime (the
blue solid line in the figure) where the associated thermal
de Broglie wavelength approaches to the Planck length
$l_{_{\rm Pl}}$.
Expanding the relation (\ref{energy}) for both low and high
temperature regimes gives
\begin{equation}\label{energy-Exp}
U=
\left\{
  \begin{array}{ll}
    \frac{3}{2}\,NT & \hspace{.5cm}\lambda
    \gg\lambda_{_{\rm Pl}},\\\\
    \frac{1}{2}\,NT &
    \hspace{.5cm}\lambda\sim\lambda_{_{\rm Pl}}.
  \end{array}
\right.
\end{equation}
The high temperature behavior of the internal energy in
Snyder model could be also more precisely understood
from the specific heat that is defined as
\begin{eqnarray}\label{S-Heat}
C_{_V}=\bigg(\frac{\partial U}{\partial T}
\bigg)_{V}=\frac{N}{2}\,\frac{2\Big(2+\frac{
\lambda^2}{\lambda_{_{\rm Pl}}^2}\Big)-
\sqrt{\pi}\,\Big(\frac{\lambda}{\lambda_{_{
\rm Pl}}}\Big)\Big(3+2\frac{\lambda^2}{
\lambda_{_{\rm Pl}}^2}\Big)\mbox{erfc}\Big[
\frac{\lambda}{\lambda_{_{\rm Pl}}}\Big]
\exp\Big[\frac{\lambda^2}{\lambda_{_{\rm Pl
}}^2}\Big]}{\bigg(2-\sqrt{\pi}\,\Big(\frac{
\lambda}{\lambda_{_{\rm Pl}}}\Big)\mbox{
erfc}\Big[\frac{\lambda}{\lambda_{_{\rm Pl
}}}\Big]\exp\Big[\frac{\lambda^2}{
\lambda_{_{\rm Pl}}^2}\Big]\bigg)^2}\,.
\end{eqnarray}
The temperature-dependent behavior of this quantity is plotted in
figure \ref{fig:2}. Expanding the specific heat (\ref{S-Heat}) for
low and high temperature regimes gives
\begin{equation}\label{S-Heat-Exp}
C_{_V}=
\left\{
  \begin{array}{ll}
    \frac{3}{2}\,N & \hspace{.5cm}\lambda
    \gg\lambda_{_{\rm Pl}},\\\\
    \frac{1}{2}\,N &
    \hspace{.5cm}\lambda\sim\lambda_{_{\rm Pl}},
  \end{array}
\right.
\end{equation}
which shows that the specific heat approaches to $C_{_V}
\rightarrow\,N/2$ at the high temperature limit (see also figure
\ref{fig:2}). As it is clear from figure \ref{fig:2}, the specific
heat is bounded as $\frac{1}{2}\leq\frac{C_{_V}}{N}\leq\frac{3}{2}$
in this setup.

\begin{figure}
\flushleft\leftskip+15em{\includegraphics[width=3in]{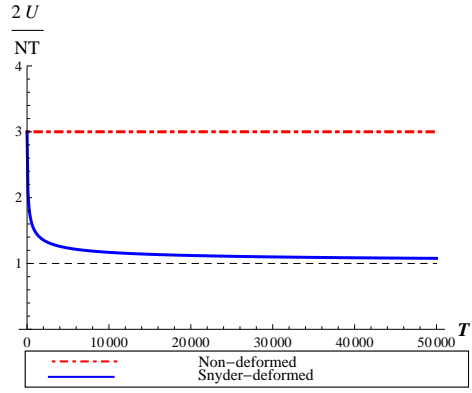}}
\hspace{3cm}\caption{\label{fig:3} Number of degrees of freedom
versus the temperature. The classical equipartition theorem of
energy states that the number of degrees of freedom for a particle
is given by $\frac{(U/N)}{(T/2)}$. Thus, as it is clear from the
figure, the number of degrees of freedom will be reduced from $3$ to
$1$ at the high temperature regime in Snyder space. This result
suggests an effective dimensional reduction of space from $3$ to $1$
dimension.}
\end{figure}

In order to understand the reduction of the number of degrees of
freedom in a more precise manner, we invoke the well-known theorem
of {\it equipartition of energy} which states that each number of
degree of freedom makes a contribution of $\frac{1}{2}\,T$ towards
the internal energy and $\frac{1}{2} $ towards the specific heat.
From the relations (\ref{energy-Exp}) and (\ref{S-Heat-Exp}), it is
clear that the number of degrees of freedom for the ideal gas
consisting of $N$ noninteracting particles which move on
three-dimensional Euclidean space ${\mathbf R}^3$ (with ${\mathbf
R}^{3N}$ configuration space), will be reduced from $3N$ to $N$ at
the high temperature regime when the thermal de Broglie wavelength
of the system becomes of the order of the Planck length
$\lambda\sim{l_{_{Pl}}}$. In other words, there is one degree of
freedom for a particle at such a high temperature regime and two
degrees of freedom will be frozen due to the quantum gravity
(minimal length) effects. See also figure \ref{fig:3} which shows
the temperature-dependent behavior of the number of degrees freedom
$\frac{(U/N)}{(T/2)}$ for a particle in Snyder model. Therefore,
according to the equipartition theorem, the number of degrees of
freedom for a particle is reduced from $3$ to $1$ at the high
temperature regime in the Snyder space. This result suggests an
effective high temperature dimensional reduction of space from $3$
to $1$ dimension in the Snyder model. Similar results have been
obtained in the context of the other phenomenological approaches to
minimal length scenario. For instance, in the context of the doubly
special relativity theories, it is shown that the total number of
degrees of freedom will be finite at the high temperature regime
which shows that all of the degrees of freedom will be frozen in
this setup \cite{Glikman}. The associated phase space then will be
compact (with finite Liouville volume) \cite{LT} and the
corresponding Hilbert space is also finite dimensional
\cite{Rovelli} (see Ref. \cite{DSR-DR} for some cosmological
consequences of such a framework). Furthermore, in the context of
polymer quantization, it is shown that the energy density of the
photon gas will be proportional to $T^{5/2}$ rather than the
standard Stephan-Boltzmann law that states the energy density will
be proportional to $T^4$ \cite{Husain}. This result also suggests an
effective reduction to $1.5$ dimensional space at the Planck scale.
Thus, the dimensional reduction at the Planck scale seems to be a
common feature of quantum gravity proposal \cite{QG-DR}.

\subsection{Pressure and Equation of State}
\begin{figure}
\flushleft\leftskip+15em{\includegraphics[width=3in]{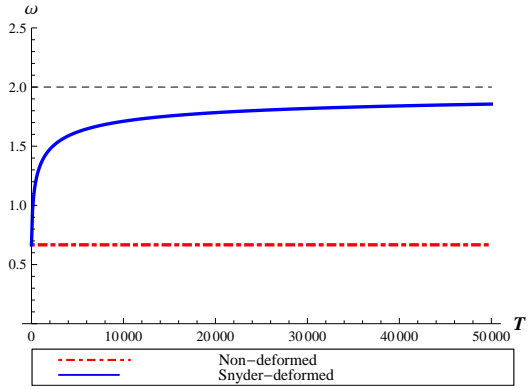}}
\hspace{3cm}\caption{\label{fig:4} Equation of state parameter
versus the temperature. While the equation of state parameter is
constant as $w=2/3$ for the standard ideal gas, it turns out to be
temperature-dependent quantity in Snyder space. As it is clear from
the figure, it approaches to $w=2$ at the very high temperature
regime when the quantum gravity (minimal length) effects dominate.
The figure shows also that $\frac{2}{3}\leq{w}\leq2$.}
\end{figure}

The thermal pressure could be obtained from the Helmholtz
energy (\ref{Helmholtz}) as
\begin{eqnarray}\label{pressure}
P=-\bigg(\frac{\partial F}{\partial V}\bigg)_{
T,N}=\frac{NT}{V}\,,
\end{eqnarray}
which shows that the equation of state relation preserves its
standard form $PV=NT$ in this setup. From the relations
(\ref{energy}) and (\ref{pressure}), the equation of state
parameter $w=\frac{P}{U/V}$ works out to be
\begin{eqnarray}\label{w}
w=\left\{\Bigg(2-\sqrt{\pi}\,\bigg(\frac{\lambda
}{\lambda_{_{\rm Pl}}}\bigg)\mbox{erfc}\bigg[
\frac{\lambda}{\lambda_{_{\rm Pl}}}\bigg]\exp
\bigg[\frac{\lambda^2}{\lambda_{_{\rm Pl}}^2}
\bigg]\Bigg)^{-1}-\frac{\lambda^2}{\lambda_{_{
\rm Pl}}^2}\right\}^{-1}\,.
\end{eqnarray}
The temperature behavior of the equation of state parameter
(\ref{w}) is plotted in figure \ref{fig:4}. The equation of state
parameter is constant as $w=\frac{2}{3}$ for the standard ideal gas.
But, as it is clear from figure \ref{fig:4}, it becomes
temperature-dependent in Snyder model at the high temperature
regime. The standard result $w=\frac{2}{3}$ can be recovered at the
low temperature regime $\lambda\gg{l}_{_{\rm Pl}}$ and also
$w\rightarrow \,2$ at the high temperature regime
$\lambda\sim{l}_{_{\rm Pl}}$. Thus, generally we have
$\frac{2}{3}\leq{w}\leq2$ which could be seen from figure
\ref{fig:4} (see also Refs. \cite{Husain} and \cite{Das} where the
same results are obtained in the contexts of polymer quantization
and holography respectively.)

\subsection{Entropy}
\begin{figure}
\flushleft\leftskip+15em{\includegraphics[width=3in]{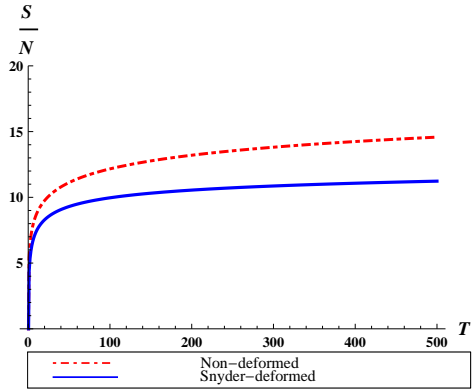}}
\hspace{3cm}\caption{\label{fig:5} Entropy versus the temperature.
It is clear that, in the Snyder model the entropy increases with a
rate smaller than the standard non-deformed case. This is because of
the fact that the number of accessible microstates is decreased at
the high temperature regime due to the quantum gravity (minimal
length) effects.}
\end{figure}

As the final thermodynamical quantity, one could obtain the
modification to the entropy from the minimal length effects. The
entropy, however, is directly related to the number of accessible
microstates and it is then natural to expect that the entropy
increases with a rate smaller than the standard non-deformed case at
the high temperature regime since we have shown that the number of
accessible microstates will be decreased at the high temperature
regime in Snyder space. The direct calculation of the entropy
justifies this claim. From the Helmholtz free energy
(\ref{Helmholtz}), the entropy of the ideal gas will be
\begin{eqnarray}\label{entropy}
\frac{S}{N}=-\left(\frac{\partial F}{\partial T}
\right)_{N,V}=1+\ln\Bigg[\frac{V}{N\lambda_{_{
\rm Pl}}^3}\Bigg(\frac{\lambda_{_{\rm Pl}}}{
\lambda}-\sqrt{\pi}\,\mbox{erfc}\bigg[\frac{
\lambda}{\lambda_{_{\rm Pl}}}\bigg]\exp\bigg[
\frac{\lambda^2}{\lambda_{_{\rm Pl}}^2}\bigg]
\Bigg)\Bigg]\nonumber\\+\Bigg(2-2\sqrt{\pi}
\bigg(\frac{\lambda}{\lambda_{_{\rm Pl}}}\bigg)
\mbox{erfc}\bigg[\frac{\lambda}{\lambda_{_{\rm
Pl}}}\bigg]\exp\bigg[\frac{\lambda^2}{
\lambda_{_{\rm Pl}}^2}\bigg]\Bigg)^{-1}-\frac{
\lambda^2}{\lambda_{_{\rm Pl}}^2}.
\end{eqnarray}
In figure \ref{fig:5}, the temperature-dependent behavior of the
ideal gas entropy (\ref{entropy}) is plotted. As it is clear from
the figure, in contrast to the non-deformed case, the entropy
increases with smaller rate in the Snyder model. This result is also
a direct consequence of the effective dimensional reduction of space
from three to one dimension for a particle in the Snyder space.

Although, as we have shown, the minimal length effects will become
important only at the very high temperature regime when the thermal
de Broglie wavelength of the system becomes of the order of the
Planck length, it is also useful to consider the low temperature
behavior in order to estimate the order of the magnitude of the
quantum gravity effects on the thermodynamical quantities of the
ideal gas. It is straightforward to show that the first order
quantum gravity corrections to all of the thermodynamical quantities
such as the internal energy (\ref{energy}), specific heat
(\ref{S-Heat}), and entropy (\ref{entropy}) are proportional to
$(\lambda_{_{\rm Pl}}/\lambda)^2$. Therefore, the thermal de Broglie
wavelength $\lambda$ is an appropriate parameter to determine when
quantum gravitational effects will become significant, much similar
in the same way as pure quantum mechanical effects become important
in the standard statistical mechanics. Indeed, the pure quantum
mechanical effects will become important when the thermal de Broglie
wavelength becomes of the order of the mean interparticle distance
$(V/N)^{1/3}$. Similarly and much in the same way, quantum
gravitational (minimal length) effects will become important when
the thermal de Broglie wavelength of the system becomes of the order
of the Planck length.

\section{Summary and Conclusions}
Existence of a universal minimal length, preferably of the order of
the Planck length, is a common address of quantum gravity candidates
such as string theory and loop quantum gravity. Beside, this issue
could be achieved from the spaces with deformed structures. In this
paper, we formulated the statistical mechanics in Snyder space in
the semiclassical regime. Existence of a minimal length, as an extra
information for the system under consideration, significantly
changes the probability distribution over the set of microstates in
this setup. We obtained the corresponding deformed invariant
Liouville volume which determines the number of accessible
microstates for a statistical system and we have found that $2/3$ of
the degrees of freedom will be frozen at the high energy regime.
Generalizing the setup into a many-particle system, we then obtained
the modified partition function for the ideal gas in the
Maxwell-Boltzmann statistics by means of the deformed Liouville
volume and we have calculated the associated thermodynamical
quantities such as the internal energy, specific heat, equation of
state parameter, and entropy in the Snyder space. The results show
that at the high temperature regime, when the thermal de Broglie
wavelength becomes of the order of the Planck length, the quantum
gravity (minimal length) effects dominate which significantly change
the thermodynamical properties of the ideal gas. Invoking the
equipartition theorem of energy, we explicitly showed that $2/3$ of
the number of degrees of freedom will be frozen at the high
temperature regime for the special case of the ideal gas which also
confirms our pervious claim. This result suggests an effective
dimensional reduction of the space from three to one dimension at
the high temperature regime which is also a common feature in
alternative approaches to quantum gravity proposal. Also, our
analysis shows that $\frac{1}{2}\leq \frac{C_{_V}}{N}
\leq\frac{3}{2}$ and $\frac{2}{3}\leq{w}\leq2$ for the specific heat
and the equation of state parameter respectively which are the
direct consequences of the effective dimensional reduction of the
space at the Planck scale. Although the quantum gravity effects
would become important only at the high temperature regime,
considering the low temperature limit is useful to estimate the
order of the magnitude of the quantum gravity corrections to the
thermodynamical quantities. Evidently, the first order corrections
to the internal energy, specific heat, and entropy are of the order
of $(\lambda_{_{\rm Pl}}/\lambda)^2$, where $\lambda_{_{\rm Pl}}$ is
the Planck scale thermal de Broglie wavelength (\ref{lambda-p}) and
$\lambda$ is the standard thermal de Broglie wavelength.

\end{document}